\documentclass[preprints,article,accept,moreauthors,10pt,a4paper]{mdpi} 
\SetLabelAlign{CenterWithParen}{\hfil({#1})\hfil}  
\usepackage{tabularx}
\usepackage{rotating}
\usepackage{mathrsfs}
\usepackage{amssymb}

\newcolumntype{C}[1]{>{\centering\arraybackslash}p{#1}}

\firstpage{1} 
\pubvolume{xx}
\issuenum{1}
\articlenumber{1}
\pubyear{2018}
\copyrightyear{2018}
\externaleditor{Academic Editor: name}
\history{Received: date; Accepted: date; Published: date}

\pdfoutput=1

\Title{Lanczos equation on light-like hypersurfaces in a cosmologically viable class of kinetic gravity braiding theories}

\title{Lanczos equation on light-like hyper surfaces in a cosmologically viable class of kinetic gravity braiding theories}
\Author{Bence Racsk\'o$^{1}$\orcidA{}; L\'aszl\'o \'A. Gergely$^{2\ast}$\orcidB{}}

\AuthorNames{Bence Racsk\'o, L\'aszl\'o \'A. Gergely}

\address{%
$^{1}$ \quad Department of Theoretical Physics, University of Szeged, Tisza L. krt. 84-86, H-6720 Szeged, Hungary; daeron806@gmail.com\\
$^{2}$ \quad Institute of Physics, University of Szeged, D\'om t\'er 9, H-6720 Szeged, Hungary; gergely@physx.u-szeged.hu}

\corres{Correspondence: laszlo.a.gergely@gmail.com}

\abstract{We discuss junction conditions across null hypersurfaces in a class of scalar-tensor gravity theories with i) second order dynamics, ii) obeying the recent constraints imposed by gravitational wave propagation, and iii) allowing for a cosmologically viable evolution. These requirements select kinetic gravity braiding models with linear kinetic term dependence and scalar field-dependent coupling to curvature. We explore a pseudo-orthonormal tetrad and its allowed gauge fixing, with one null vector standing as the normal, the other being transversal to the hypersurface. We derive a generalization of the Lanczos equation in a 2+1 decomposed form, relating the energy density, current and isotropic pressure of a distributional source to the jumps in the transverse curvature and transverse derivative of the scalar. Additionally we discuss a scalar junction condition and its implications for the distributional source. }


\begin{document}


\section{Introduction}

Scalar-tensor gravity theories give viable modifications of general
relativity, in which accelerated expansion could be recovered without dark
energy at late times; well-tested Solar System constraints could be obeyed
(for example through the Vainshtein mechanism); and the recent constraint
from gravitational wave detections \cite{GW1,GW2,GW3,GW4,GW5,GW6,GW7-11} on
the propagation speed of the tensorial modes could be successfully
implemented. Indeed from the class of Hordeski theories ensuring second
order dynamics for both the scalar field and the metric tensor \cite%
{Horndeski, Deffayet} a subclass has been identified \cite%
{GWc1,GWc2,GWc3,GWc4}, in which gravitational waves propagate with the speed
of light (as verified both from the almost coincident detection with
accompanying $\gamma $-rays in the case of the neutron star binary merger
and from a strigent test of the dispersion relations disruling massive modes
for the 10 black hole mergers). This subclass contains cubic derivative
couplings of the scalar field in the Lagrangian, known as kinetic gravity
braiding \cite{KineticBraidingKT,KineticBraidingDPSV}. In the Jordan frame
the curvature couples with the scalar through an unspecified function of the
scalar field.

This class of scalar-tensor gravity models could be further restricted by
the requirement to ensure a viable cosmological evolution. In Ref. \cite{KTF}
it has been proven that for a kinetic gravity braiding model with Lagrangian
only linearly and quadratically depending on the kinetic term $X=-\left(
\nabla \phi \right) ^{2}/2$ an autonomous system of equations governs the
dynamics, leading to a number of fixed points for the background dynamics,
three of them representing consecutive radiation, matter and dark energy
dominated regimes (see for example Fig. 1. of Ref. \cite{KTF}). The same
model was further analysed from the string theory motivated point of view of
avoiding de Sitter regimes, which are not embeddable in string theory \cite%
{HBBR}. Cross-correlating this model class with the requirement of the
propagation of tensorial modes with the speed of light, the quadratic
dependence has to be dropped. In this paper we consider this class of
kinetic gravity braiding models with only linear dependence on the kinetic
terms and analyze the junctions across null hypersurfaces.

Junction conditions in general relativity are known either for spatial or
temporal hypersurfaces \cite{Israel} or for null hypersurfaces \cite{BI}, 
\cite{Poisson}. The latter are more sophisticated, as the normal to the
hypersurface is not suitable for a 3+1 space-time decomposition, being in
the same time also tangent. The decomposition can be done with respect to a
transverse vector, with the gauge arising from its nonunique choice dropping
out from the final results \cite{BI}, or by employing a pseudo-orthonormal
basis with two null vectors, one of them playing the role of the normal, the
other being transversal \cite{Poisson}. The distributional contribution
arising in the curvature from the possible discontinuity of the metric
derivative across the hypersurface is related to singular sources on the
hypersurface through the Lanczos equation. The same technique led to the
derivation of the dynamics on a brane embedded in a 5-dimensional bulk \cite%
{SMS,Fried,VarFried}.

In the full Horndeski class of scalar-tensor gravity theories junction
conditions across spatial or temporal hypersurfaces have been derived \cite%
{PadillaS,NK}, but the null case stays uncovered, although its importance is
undoubtful as all electromagnetic and gravitational shock-waves propagate
along such hypersurfaces.

Here we propose to derive such junction conditions for the class of kinetic
gravity braiding theories with linear kinetic term, which as discussed above
are both cosmologically viable and obey the gravitational wave constraints.
This generalizes our earlier work on null junctions in Brans-Dicke theories 
\cite{universe}.

The notations are as follows: space-time indices are greek, 2-dimensional
spatial indices are latin capital letters. The soldering of any quantity $A$, with values $A^+$ and $A^-$ on the two sides of the hypersurface is $\tilde{A}%
=A^{+}\Theta \left( f\right) +A^{-}\Theta \left( -f\right) $, where $\Theta $
is the step function. The average on the hypersurface is denoted as $%
\left\langle A\right\rangle =\left( A^{+}+A^{-}\right) /2$, the jump over
the hypersurface as $\left[ A\right] =A^{+}-A^{-}$.



\section{Equations of motion}

The assumed Lagrangian%
\begin{equation}
L_{GKGB}=\underbrace{B(\phi )X+V(\phi )}_{L_{2}}\underbrace{-2\xi (\phi
)\square \phi X}_{L_{3}}\underbrace{+\frac{1}{2}F(\phi )R}_{L_{4}}
\end{equation}%
with $B,\xi ,F$ arbitrary functions of the scalar field yields the following
expressions through the variation of metric%
\begin{align}
E_{\mu \nu }^{(2)}& =-\frac{1}{2}B(\phi )\left( Xg_{\mu \nu }-\phi _{\mu
}\phi _{\nu }\right) -\frac{1}{2}V(\phi )g_{\mu \nu }, \\
E_{\mu \nu }^{(3)}& =\xi (\phi )\square \phi \phi _{\mu }\phi _{\nu }+2\xi
^{\prime }(\phi )X\left( \phi _{\mu }\phi _{\nu }+Xg_{\mu \nu }\right) +2\xi
(\phi )X_{(\mu }\phi _{\nu )}-\xi (\phi )X_{\kappa }\phi ^{\kappa }g_{\mu
\nu }, \\
E_{\mu \nu }^{(4)}& =\frac{1}{2}\left\{ F(\phi )G_{\mu \nu }+\left(
F^{\prime }(\phi )\square \phi -2F^{\prime \prime }(\phi )X\right) g_{\mu
\nu }-F^{\prime }(\phi )\phi _{\mu \nu }-F^{\prime \prime }(\phi )\phi _{\mu
}\phi _{\nu }\right\} ,
\end{align}%
and through the variation of the scalar field%
\begin{align}
E_{\phi }^{(2)}& =B(\phi )\square \phi -B^{\prime }(\phi )X+V^{\prime }(\phi
), \\
E_{\phi }^{(3)}& =\xi (\phi )\left\{ (\square \phi )^{2}-\phi _{\mu \nu
}\phi ^{\mu \nu }-R^{\mu \nu }\phi _{\mu }\phi _{\nu }\right\} -2\xi
^{\prime \prime }(\phi )X^{2}, \\
E_{\phi }^{(4)}& =\frac{1}{2}F^{\prime }(\phi )R,
\end{align}%
where $\phi _{\mu }\equiv \nabla _{\mu }\phi $ and $\phi _{\mu \nu }\equiv
\nabla _{\nu }\nabla _{\mu }\phi $. The Ricci curvature tensor appears in
the expression $E_{\phi }^{(3)}$ through the Ricci identity $[\nabla _{\mu
},\nabla _{\nu }]V^{\kappa }=R^{\kappa }{}_{\lambda \mu \nu }V^{\lambda }$,
which has been used to get rid of \textit{third derivatives} of $\phi $.

These are the left hand sides of the equations of motion (EoMs). The right
hand sides are half of the energy-momentum tensor for the metric variation
of the matter action and zero for the scalar field variation as in the
Jordan frame the matter does not couple to the scalar field.

\section{Junction conditions}

\subsection{The extrinsic formulation}

We employ a pseudo-orthonormal basis with two null vectors $N^{\mu }$ and $%
L^{\mu }$, the first of them being the normal (surface gradient, which is
also tangent) to the hypersurface $\Sigma $, the other playing the role of
the transverse vector, with respect to which we perform a (2+1)+1
decomposition \cite{Poisson}. The normalization is $L^{\mu }N_{\mu }=-1$.
The continuity of both the metric tensor $g_{\mu \nu }$ and scalar $\phi $
are imposed over the hypersurface: $\left[ \phi \right] =\left[ g_{\mu \nu }%
\right] =0$. Their first derivatives in the null transverse direction $\phi
_{L}\equiv L^{\mu }\partial _{\mu }\phi $ and $L^{\rho }\partial _{\rho
}g_{\mu \nu }$ may have a jump%
\begin{equation}
\zeta =\left[ \phi _{L}\right] ,\quad c_{\mu \nu }=\left[ L^{\rho }\partial
_{\rho }g_{\mu \nu }\right] ,
\end{equation}%
and since all tangential derivatives are assumed to be continuous, we have 
\begin{equation}
\lbrack \phi _{\mu }]=-N_{\mu }\zeta ,\quad \lbrack \partial _{\kappa
}g_{\mu \nu }]=-N_{\kappa }c_{\mu \nu }.
\end{equation}%
The second order derivatives appearing in the equations of motion%
\begin{align}
E_{\mu \nu }& \equiv \tilde{E}_{\mu \nu }+\mathscr{E}_{\mu \nu }\delta
\left( f\right) =\frac{1}{2}\left( \tilde{T}_{\mu \nu }+\mathscr{T}_{\mu \nu
}\delta \left( f\right) \right) , \\
E_{\phi }& \equiv \tilde{E_{\phi }}+\mathscr{E}_{\phi }\delta \left(
f\right) =0.
\end{align}%
lead to the distributional contributions $\mathscr{E}_{\mu \nu }$ and $%
\mathscr{E}_{\phi }$\ along the thin shell, arising from the derivative of
the step function. All quantities with tilde are the regular contributions
to the respective quantities. For consistency we also include a
distributional energy-momentum tensor $\mathscr{T}_{\mu\nu}$ together with the regular one $\tilde{T}_{\mu\nu}$. In the argument
of the delta distribution, $f$ denotes a function which generates the
hypersurface as its zero set. For convenience we also assume that $N_{\mu
}=\nabla _{\mu }f$.

We introduce the notations 
\begin{equation}
c_{\mu }=c_{\mu \nu }N^{\nu },\quad c^{\dagger }=c_{\mu }N^{\mu },\quad
c=c_{\mu }^{\mu },
\end{equation}%
and give explicitly the jump of the connection as 
\begin{equation}
\lbrack \Gamma _{\mu \nu }^{\kappa }]=-\frac{1}{2}\left( N_{\mu }c_{\nu
}^{\kappa }+N_{\nu }c_{\mu }^{\kappa }-N^{\kappa }c_{\mu \nu }\right) ,
\end{equation}%
hence the singular parts of the curvature tensor and its traces become 
\begin{align}
\mathscr{R}^{\kappa }{}_{\lambda \mu \nu }& =-\frac{1}{2}\left( N_{\mu
}c_{\nu }^{\kappa }N_{\lambda }-N_{\nu }c_{\mu }^{\kappa }N_{\lambda
}+N_{\nu }c_{\mu \lambda }N^{\kappa }-N_{\mu }c_{\nu \lambda }N^{\kappa
}\right) \\
\mathscr{R}_{\mu \nu }& =-\frac{1}{2}\left( N_{\mu }c_{\nu }+N_{\nu }c_{\mu
}-N_{\mu }N_{\nu }c\right) \\
\mathscr{R}& =-c^{\dagger }.
\end{align}%
In particular, the singular part of the Einstein tensor is 
\begin{equation}
\mathscr{G}_{\mu \nu }=-\frac{1}{2}\left( N_{\mu }c_{\nu }+N_{\nu }c_{\mu
}-N_{\mu }N_{\nu }c-c^{\dagger }g_{\mu \nu }\right) .
\end{equation}%
We also give the jumps and singular parts of the quantities constructed from
the scalar field. As a calligrafic version of $\phi $ is not
catchy, in the decomposition $A=\tilde{A}+\mathscr{A}\delta (f)$%
  we introduce the alternative notation $\mathscr{A}\equiv\text{Sing}(A) $, denoting the
singular part of the arbitrary quantity $A$.

For the scalar field, we have 
\begin{equation}
\text{Sing}(\phi _{\mu \nu })=-\zeta N_{\mu }N_{\nu },\quad \text{Sing}%
(\square \phi )=-\zeta N_{\mu }N^{\mu }=0,\quad \lbrack X]=\phi _{N}\zeta ,
\end{equation}%
where $\phi _{N}=N^{\mu }\phi _{\mu }$ is the normal derivative. We note
that the value of $\phi _{N}$ on the hypersurface is unambigous, being a
tangential derivative, which is continuous.

Explicit calculation gives the hypersurface contributions to the left hand
side of the tensorial EoMs: 
\begin{align}
\mathscr{E}_{\mu \nu }^{(2)}& =0 \\
\mathscr{E}_{\mu \nu }^{(3)}& =\xi (\phi )\zeta \left( 2\phi _{N}N_{(\mu
}\langle \phi _{\nu )}\rangle -\phi _{N}^{2}g_{\mu \nu }\right)
\label{extensor3} \\
\mathscr{E}_{\mu \nu }^{(4)}& =\frac{1}{2}\left( F(\phi )\mathscr{G}_{\mu
\nu }+F^{\prime }(\phi )\zeta N_{\mu }N_{\nu }\right)  \label{extensor4}
\end{align}%
and the hypersurface contributions to the left hand side of the scalar EoMs: 
\begin{align}
\mathscr{E}_{\phi }^{(2)}& =0 \\
\mathscr{E}_{\phi }^{(3)}& =\xi (\phi )\left( 2\zeta N^{\mu }N^{\nu }\langle
\phi _{\mu \nu }\rangle +\phi _{N}c^{\mu }\langle \phi _{\mu }\rangle -\frac{%
1}{2}\phi _{N}^{2}c\right)  \label{exscalar3} \\
\mathscr{E}_{\phi }^{(4)}& =-\frac{1}{2}F^{\prime }(\phi )c^{\dagger }
\label{exscalar4}
\end{align}

\subsection{The intrinsic formulation}

The above equations are expressed in a four dimensional coordinate system
smooth across the hypersurface. Such coordinate systems may be difficult to
construct, hence it would be more practical to use coordinantes intrinsic to the
junction hypersurface.

The hypersurface contributions to the left hand side of the tensor EoMs (\ref%
{extensor3},\ref{extensor4}) are tangential in the sense that 
\begin{equation}
\mathscr{E}_{\mu \nu }^{(3)}N^{\nu }=\mathscr{E}_{\mu \nu }^{(4)}N^{\nu }=0,
\end{equation}%
hence we may expand them in a basis adapted to the junction hypersurface $%
\Sigma $. We choose this basis as $\left( L^{\mu },N^{\mu },e_{2}^{\mu
},e_{3}^{\mu }\right) $, where the $e_{A}^{\mu }$ are two spacelike tangent
vector fields to $\Sigma $, satisfying 
\begin{equation}
N_{\mu }e_{A}^{\mu }=L_{\mu }e_{A}^{\mu }=0.
\end{equation}%
For a fixed choice of $N^{\mu }$, we may always choose $e_{A}^{\mu }$ such
that the vector fields $(N^{\mu },e_{2}^{\mu },e_{3}^{\mu })$ form a \textit{%
holonomic} set, but this is not imperative (we may also choose them
to form a pseudo-orthonormal system). The following statements are valid
also in the anholonomic case. The inner products of the spacelike vectors generate a spacelike induced metric 
\begin{equation}
q_{AB}=g_{\mu \nu }e_{A}^{\mu }e_{B}^{\nu }
\end{equation}%
on the two-dimensional subspaces
spanned by the vectors $e_{A}^{\mu }$. Its inverse is denoted $q^{AB}$
(capital latin indices are raised and lowered by either the metric or its
inverse). The completeness relation of the adapted basis is 
\begin{equation}
g^{\mu \nu }=-L^{\mu }N^{\nu }-N^{\mu }L^{\nu }+q^{AB}e_{A}^{\mu }e_{B}^{\nu
}.
\end{equation}

We further denote $e_{1}^{\mu }=N^{\mu }$, with the latin indices $a,b,...$
taking the values $1,2,3$. The extrinsic curvature $K_{ab}=e_{a}^{\mu
}e_{b}^{\nu }\frac{1}{2}\mathcal{L}_{N}g_{\mu \nu }$ is unsuitable to
describe the transversal change in the metric, as $N^{\mu }$ is also
tangential. For this reason we introduce the \textit{transverse curvature} 
\cite{Poisson}: 
\begin{equation}
\mathcal{K}_{ab}=\frac{1}{2}e_{a}^{\mu }e_{b}^{\nu }\mathcal{L}_{L}g_{\mu
\nu },
\end{equation}%
with its jump related to $c_{\mu \nu }$ as 
\begin{equation}
\lbrack \mathcal{K}_{ab}]=\frac{1}{2}e_{a}^{\mu }e_{b}^{\nu }c_{\mu \nu }.
\end{equation}%
The singular part (hypersurface contribution) of the Einstein equation is
but the generalized Lanczos equation 
\begin{equation}
\mathscr{E}^{\mu \nu }=\frac{1}{2}\mathscr{T}^{\mu \nu },  \label{Lanczos}
\end{equation}%
where $\mathscr{E}$ is the sum of the terms (\ref{extensor3},\ref{extensor4}%
). As the left hand side is purely tangential, the distributional
stress-energy-momentum tensor admits the decomposition 
\begin{equation}
\mathscr{T}^{\mu \nu }=\rho N^{\mu }N^{\nu }+j^{A}\left( N^{\mu }e_{A}^{\nu
}+e_{A}^{\mu }N^{\nu }\right) +p^{AB}e_{A}^{\mu }e_{B}^{\nu },
\label{semdecomp}
\end{equation}%
where $\rho ,j^{A}$ and $p^{AB}$ are the energy density, current vector and
stress tensor of the distributional source. These quantities, defined as the
components emerging with respect to the intrinsic triad of vectors, can be
evaluated even when the bulk coordinates do not match smoothly along $\Sigma 
$. They are defined as 
\begin{equation}
\rho =2\mathscr{E}_{\mu \nu }L^{\mu }L^{\nu },\quad j_{A}=-2\mathscr{E}_{\mu
\nu }L^{\mu }e_{A}^{\nu },\quad p^{AB}=2\mathscr{E}_{\mu \nu }e_{A}^{\mu
}e_{B}^{\nu }.
\end{equation}

The 2+1 decomposition of Eq. (\ref{Lanczos}) yields an isotropic pressure $%
p^{AB}=pq^{AB}$ and 
\begin{align}
\rho & =F(\phi )[\mathcal{K}_{AB}]q^{AB}+F^{\prime }(\phi )[\phi _{L}]-2\xi
(\phi )\phi _{N}[\phi _{L}^{2}],  \label{rhoeq} \\
j_{A}& =-F(\phi )[\mathcal{K}_{NA}]+2\xi (\phi )[\phi _{L}]\phi _{N}\phi
_{A},  \label{jeq} \\
p& =F(\phi )[\mathcal{K}_{NN}]-2\xi (\phi )[\phi _{L}]\phi _{N}^{2},
\label{peq}
\end{align}%
where $\phi _{A}=e_{A}^{\mu }\phi _{\mu }$, $\mathcal{K}_{NA}\equiv \mathcal{%
K}_{1A}$ and $\mathcal{K}_{NN}\equiv \mathcal{K}_{11}$.

The scalar equation is 
\begin{eqnarray}
0 &=&\xi (\phi )\phi _{N}^{2}q^{AB}[\mathcal{K}_{AB}]-2\xi (\phi )\phi
_{N}\phi ^{A}[\mathcal{K}_{NA}]  \notag \\
&&+\left( F^{\prime }(\phi )+2\xi (\phi )\phi _{N}\langle \phi _{L}\rangle
\right) [\mathcal{K}_{NN}]-2\xi (\phi )[\phi _{L}]\left( \phi _{NN}-\langle 
\mathcal{K}_{NN}\rangle \phi _{N}\right) ,
\end{eqnarray}%
which contains jumps and averages. However by exploring the relation $%
[A]\langle B\rangle +\langle A\rangle \lbrack B]=[AB]$ the averages can be
transformed away to obtain 
\begin{align}
0& =\xi (\phi )\phi _{N}^{2}q^{AB}[\mathcal{K}_{AB}]-2\xi (\phi )\phi
_{N}\phi ^{A}[\mathcal{K}_{NA}]+F^{\prime }(\phi )[\mathcal{K}_{NN}]  \notag
\\
& -2\xi (\phi )[\phi _{L}]\phi _{NN}+2\xi (\phi )\phi _{N}[\phi _{L}\mathcal{%
K}_{NN}]  \label{scaleq}
\end{align}%
\bigskip 

Equations (\ref{rhoeq}-\ref{peq}) provide generalisations of the Lanczos
equation and equation (\ref{scaleq}) a constraint on the distributional
sources.

\subsection{Gauge fixing}

At this point it is worthwhile to remember that there is still gauge freedom
in the tetrad choice. The normal vector field is autoparallel \cite{Poisson} 
\begin{equation}
N^{\nu }\nabla _{\nu }N^{\mu }=\kappa N^{\mu }
\end{equation}%
with the \textit{non-affinity parameter} $\kappa =\mathcal{K}_{NN}$. If the
null fields are rescaled as $\bar{N}^{\mu }=e^{\alpha }N^{\mu }$ and $\bar{L}%
^{\mu }=e^{-\alpha }L^{\mu }$ with some function $\alpha $ defined on the
hypersurface, then the non-affinity parameter changes as 
\begin{equation}
\bar{\kappa}=e^{\alpha }\left( N^{\nu }\nabla _{\nu }\alpha +\kappa \right) ,
\end{equation}%
while 
\begin{equation*}
\phi _{\bar{L}}=e^{-\alpha }\phi _{L}
\end{equation*}%
hence%
\begin{equation}
\phi _{\bar{L}}\mathcal{\bar{K}}_{\bar{N}\bar{N}}=\phi _{L}\left( N^{\nu
}\nabla _{\nu }\alpha +\mathcal{K}_{NN}\right) .
\end{equation}%
It is possible to achieve 
\begin{equation}
\lbrack \phi _{\bar{L}}\mathcal{\bar{K}}_{\bar{N}\bar{N}}]=0
\end{equation}%
through any solution of the differential equation 
\begin{equation}
\frac{\partial \alpha }{\partial \lambda }=-\langle \kappa \rangle -\frac{%
\langle \phi _{L}\rangle }{\left[ \phi _{L}\right] }[\kappa ],
\end{equation}%
where $\lambda $ is a coordinate adapted to $N^{\mu }$ and the ratio $%
\langle \phi _{L}\rangle /\left[ \phi _{L}\right] $ is a function on the
hypersurface, being evaluated there. Hence, in this gauge, the last term of
eq. (\ref{scaleq}) drops out.

\section{Discussion of the junction conditions}

From the 2+1 decomposed form of the tensorial junction conditions we may
express the jumps in the components of the transverse curvature in terms of
the distributional energy density, current and isotropic pressure, also the
jump of the transverse derivative of the scalar field and in its square, as
follows

\begin{equation}
\lbrack \mathcal{K}_{AB}]q^{AB}=\frac{\rho }{F}-\left( \ln F\right) ^{\prime
}[\phi _{L}]+\frac{2\xi \phi _{N}}{F}[\phi _{L}^{2}],  \label{t1}
\end{equation}%
\begin{equation}
\lbrack \mathcal{K}_{NA}]=-\frac{j_{A}}{F}+\frac{2\xi \phi _{N}}{F}\phi
_{A}[\phi _{L}],  \label{t2}
\end{equation}%
\begin{equation}
\lbrack \mathcal{K}_{NN}]=\frac{p}{F}+\frac{2\xi \phi _{N}^{2}}{F}[\phi
_{L}],  \label{t3}
\end{equation}%
Then the scalar junction equation (in the gauge where $[\phi _{L}\mathcal{K}%
_{NN}]=0$) becomes 
\begin{eqnarray}
&&F^{\prime }p+\xi \phi _{N}\left( \phi _{N}\rho +2\phi ^{A}j_{A}\right)  
\notag \\
&=&\xi \left( 2F\phi _{NN}-F^{\prime }\phi _{N}^{2}+4\xi \phi _{N}^{2}\phi
_{A}\phi ^{A}\right) [\phi _{L}]-2\xi ^{2}\phi _{N}^{3}[\phi _{L}^{2}].
\label{sc}
\end{eqnarray}%
There are two cases when these equations simplify considerably: A) when
there is no cubic derivative coupling $\xi =0$, and B) when the normal
derivative of the scalar field vanishes $\phi _{N}=0$. In both cases the
scalar equation (\ref{sc}) shows that there is no isotropic pressure $p=0$,
the third Lanczos equation (\ref{t3}) implies $[\mathcal{K}_{NN}]=0,$ the
second Lanczos equation (\ref{t2}) gives the current as $j_{A}=-F\left( \phi
\right) [\mathcal{K}_{NA}{}]$, finally the first Lanczos equation (\ref{t1})
constraints the energy density as $\rho =F\left( \phi \right) [\mathcal{K}%
_{AB}]q^{AB}+F^{\prime }\left( \phi \right) [\phi _{L}]$.

\section{Concluding remarks}

By exploring a formalism based on a transverse null vector to the null
hypersurface we derived junction conditions across null shells in the
kinetic gravity braiding theories with linear kinetic term dependence, in
which the curvature and the scalar couples through a generic scalar field
dependent function. These scalar-tensor theories obey both the gravitational
wave constraints and could exhibit a viable cosmological evolution through
radiation, matter and dark energy dominated fixed points. Our formalism
gives the necessary equations to discuss energetic shock waves propagating
with the speed of light in these models.

The junction conditions contain the 2+1 decomposed form of the tensorial
equation, a generalization of the general relativistic Lanczos equation.
This relates the jump in the transverse curvature to the distributional
energy density, current and isotropic pressure. In the relations also the
jump of the transverse derivative of the scalar and its square are involved.
An additional scalar equation, without counterpart in general relativity,
constraints all these functions.    

If either there is no cubic derivative coupling terms $\xi =0$, or the
scalar field does not change in the normal direction to the null
hypersurface $\phi _{N}=0$, the junction conditions simplify considerably,
leaving the possibility of a distributional source without pressure 
\begin{equation}
\mathscr{T}^{\mu \nu }=\left( F\left( \phi \right) [\mathcal{K}%
_{AB}]q^{AB}+F^{\prime }\left( \phi \right) [\phi _{L}]\right) N^{\mu
}N^{\nu }-F\left( \phi \right) [\mathcal{K}_{NA}{}](N^{\mu }e_{A}^{\nu
}+e_{A}^{\mu }N^{\nu }),
\end{equation}%
together with the geometric condition $[\mathcal{K}_{NN}]=0$. These
generalize the corresponding result found for Brans-Dicke theories in Jordan
frame \cite{universe}.

\acknowledgments{This work was supported by the Hungarian National Research Development and Innovation Office (NKFIH) in the form of the grant 123996 and carried out in the framework of COST actions CA15117 (CANTATA) and CA16104 (GWverse), supported by COST
(European Cooperation in Science and Technology). During the preparation of
this manuscript B.R. was supported by the UNKP-18-3 New National Excellence
Program of the Ministry of Human Capacities of Hungary.}




\end{document}